\documentclass[journal,twoside,web]{ieeetran}

\usepackage{amsmath,amssymb,amsfonts}
\usepackage{graphicx}
\usepackage{textcomp}
\usepackage{epstopdf}
\usepackage{siunitx}
\usepackage{amssymb}
\usepackage{subfig}

\usepackage{hyperref}
\hypersetup{
    colorlinks=true,
    linkcolor=black,
    urlcolor=blue
    }

\begin{document}
\title{On MOSFET Threshold Voltage Extraction Over the Full Range of Drain Voltage Based on Gm/ID}
\author{Nikolaos~Makris and Matthias~Bucher, \IEEEmembership{Member, IEEE}
%	, \IEEEmembership{Member, IEEE}
\thanks{This work was partly supported under Project INNOVATION-EL-Crete (MIS 5002772). The review of this article was arranged by Editor  . \textit{Corresponding author: Matthias Bucher.}}
\thanks{N. Makris is with the School of Electrical and Computer Engineering, Technical University of Crete, GR-73100 Chania, Greece and also with the Foundation for Research and Technology-Hellas (FORTH), GR-71110 Heraklion, Greece. M. Bucher is with the School of Electrical and Computer Engineering, Technical University of Crete, GR-73100 Chania, Greece (e-mail: bucher@electronics.tuc.gr).}
}
\maketitle

\begin{abstract}
A MOSFET threshold voltage extraction method covering the entire range of drain-to-source voltage, from linear to saturation modes, is presented. Transconductance-to-current ratio is obtained from MOSFET transfer characteristics measured at low to high drain voltage. Based on the charge-based modeling approach, a near-constant value of threshold voltage is obtained over the whole range of drain voltage for ideal, long-channel MOSFETs. The method reveals a distinct increase of threshold voltage versus drain voltage for halo-implanted MOSFETs in the low drain voltage range. The method benefits from moderate inversion operation, where high-field effects, such as vertical field mobility reduction and series resistances, are minimal. The present method is applicable over the full range of drain voltage, is fully analytical, easy to be implemented, and provides more consistent results when compared to existing methods.

\end{abstract}
\section{Introduction}

\IEEEPARstart{T}hreshold voltage extraction methods for MOSFETs that cover the full range of drain voltage are rare, mainly due to the regional approaches underlying many conventional methods \cite{vtext1}. Moderate inversion in MOSFETs, where diffusion and drift transport components have comparable magnitude, covers an important range of current. The recognition that $G_m/I_D$ has a quite universal behavior over technology, bias ranges (mostly in saturation), and temperature \cite{gmi1}, \cite{gmi2}, enhances the interest in related techniques. The transconductance-to-current ratio change (cTCR) methods relate the extremum of $\partial(G_m/I_D)/\partial V_G$ to threshold voltage in \cite{dgmi1995}, \cite{dgmi2010}, \cite{dgmi2011} and may be applied from linear to saturation operation. A drawback of these methods is the increased noise due to the necessary successive derivations. Methods relying directly on transconductance-to-current ratio \cite{gmi_cunha}, \cite{gmi_siebel} emphasize on linear mode operation. The adjusted constant-current (ACC) \cite{acc} and generalized constant-current (GCC) \cite{gcc} techniques relate a threshold current criterion directly to $G_m/I_D$. The ACC method is applicable over the full range of drain voltage \cite{acc}. All the aforementioned methods relate to $G_m/I_D$ and hence share a sound physical basis. None of them, to the best of our knowledge, has been demonstrated in detail over the full range of drain voltage.

A novel, simple, and fully analytic method of threshold voltage extraction based on transconductance-to-current ratio (TCR) is introduced in this work. The latter is obtained from $I_D-V_G$ measured at different values of $V_{DS}$. The method is based on the theory of the charge-based model \cite{invq}, \cite{ekvbook}. While formally equivalent to the ACC \cite{acc} method, the usage of the Lambert W function renders the procedure analytically explicit. The method is among the few with which threshold voltage may be consistently obtained over the whole range of drain voltage, from linear mode to saturation. The method is shown to yield a threshold voltage independent of drain voltage for ideal long-channel MOSFETs. For halo-doped MOSFETs, the method reveals a characteristic increase of threshold voltage at $V_{DS} < 4\cdot U_T$. The method is particularly suited for the investigation of DIBL effects including long-channel DIBL \cite{ldibl}.

\section{MOSFET Threshold Voltage Extraction Method for Full Drain Voltage Range}
The charge-voltage relation in a long-channel MOSFET is given in \cite{invq} as,
\begin{equation}
\begin{split}
      v_p - v_{d,s} =  2q_{d,s} + \ln (  q_{d,s} ) 
      \label{mosvq}
\end{split}
\end{equation}
where $v_{p,g,s,d} = V_{P,G,S,D}/U_T$ are the pinch-off, gate, source and drain voltages (all referred to local substrate), and thermal voltage is $U_T = kT/q$. The pinch-off voltage $V_P$ is estimated as in \cite{acc},
\begin{equation}
\begin{split}
      V_P \approx (V_G-V_T)/n
      \label{vpvg}
\end{split}
\end{equation}
where $V_T$ and $n$ is the threshold voltage and slope factor, respectively. $q_{s,d}$ are the normalized mobile charge densities at source and drain sides \cite{acc}, respectively. Solving for mobile charges results in,
\begin{figure*}[ht!]
	\centerline{%
		\begin{tabular}{ccc}
	   {\includegraphics[scale=0.115]{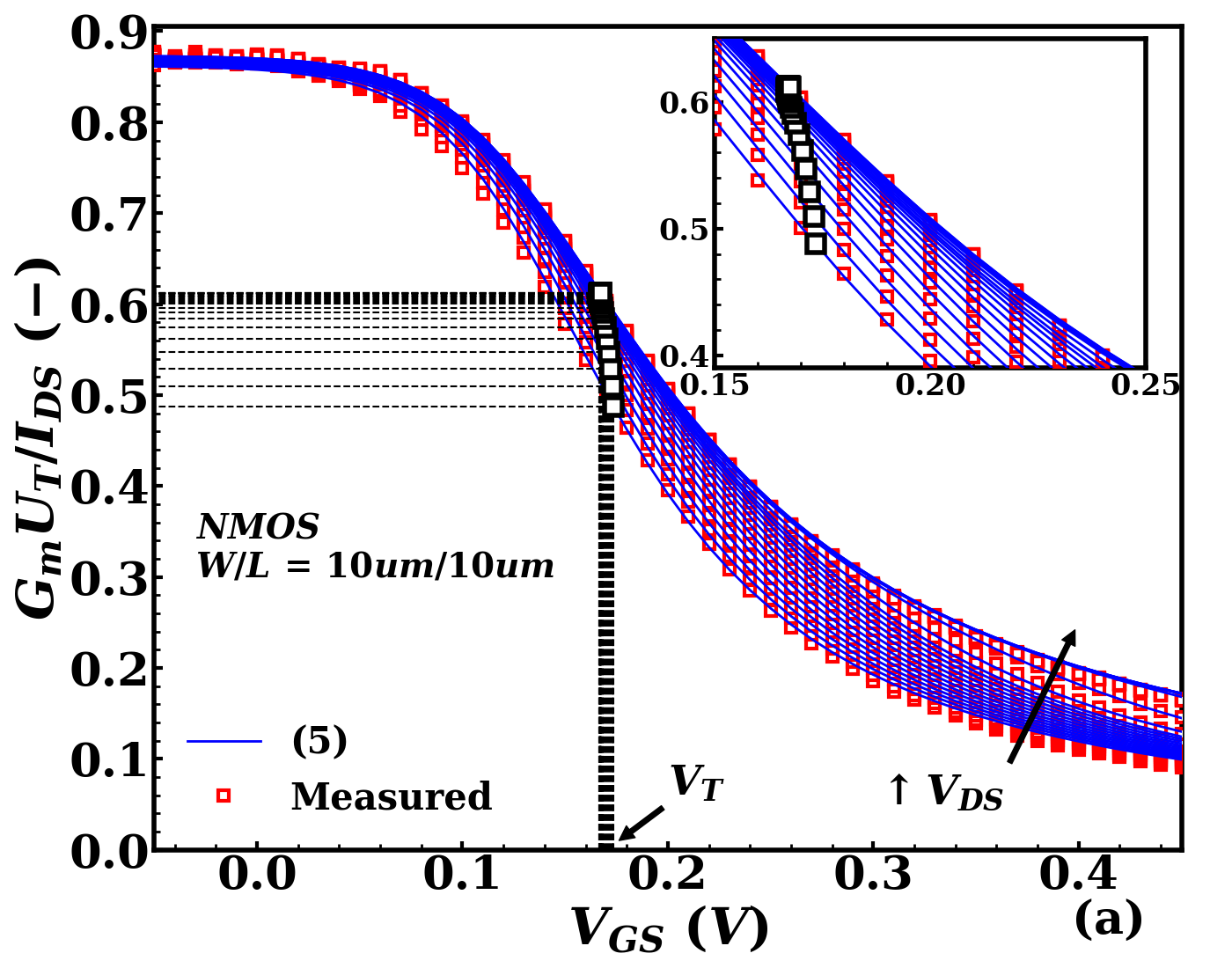}} & 
	    {\includegraphics[scale=0.115]{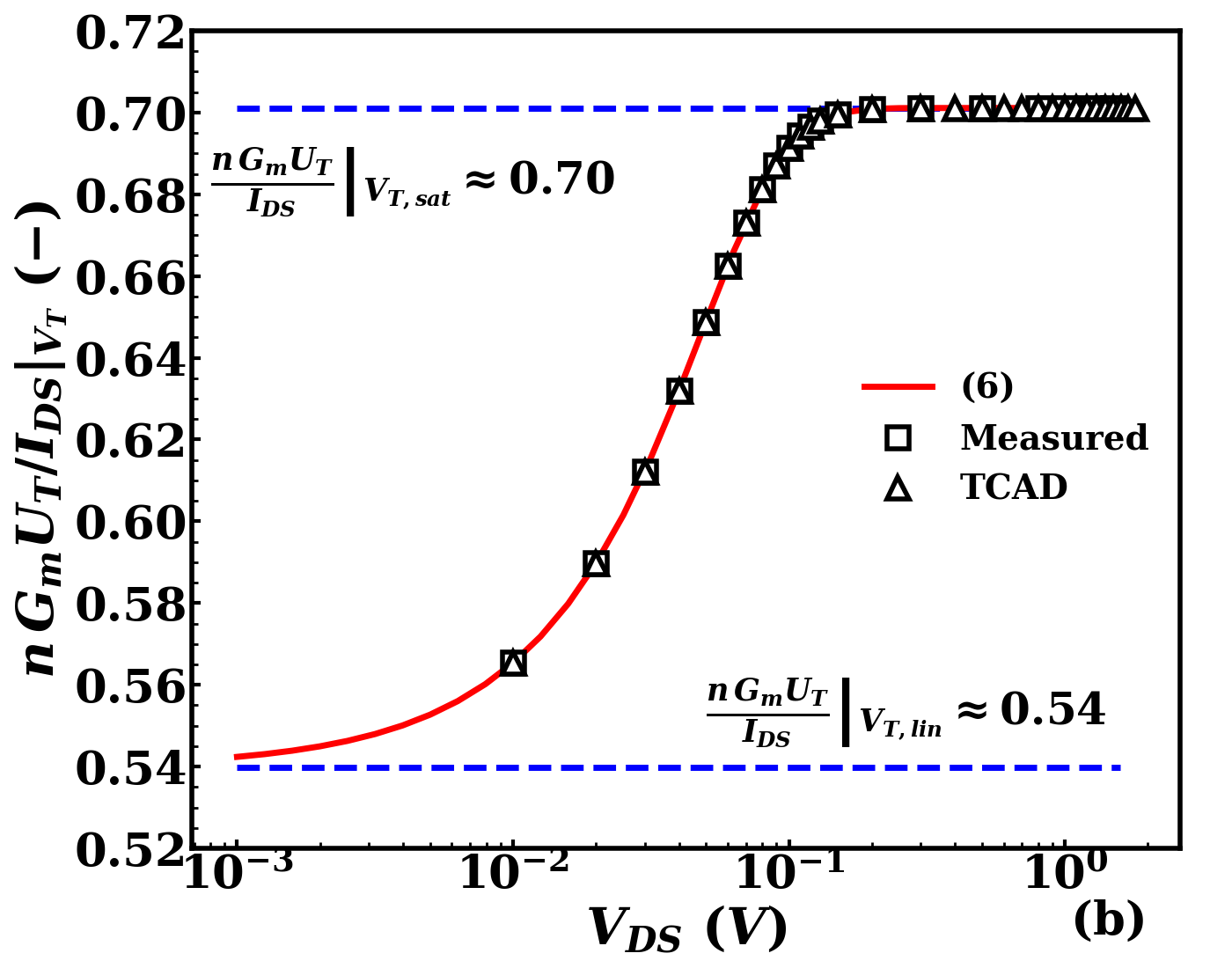}}  &
		{\includegraphics[scale=0.115]{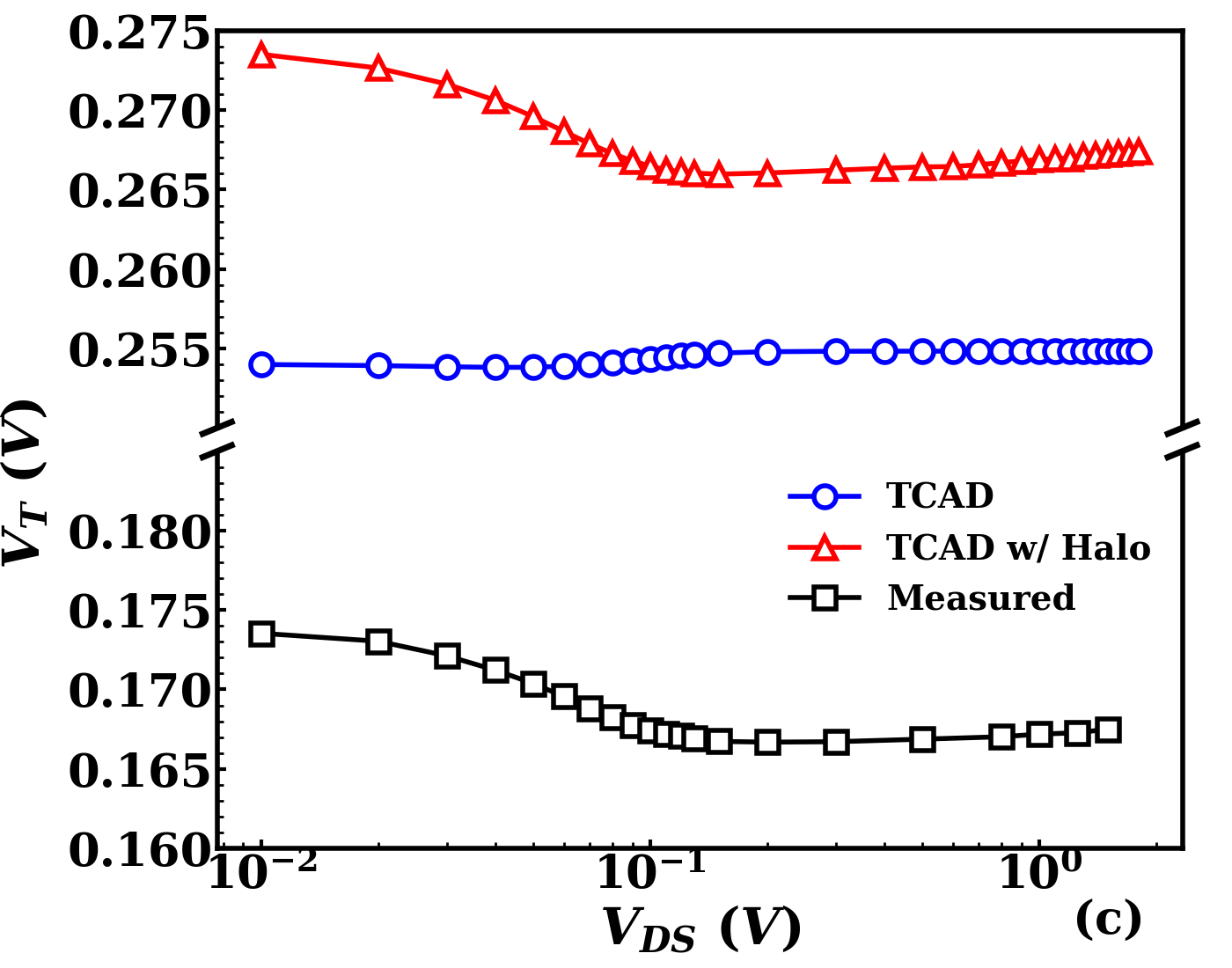}} 
	\end{tabular}} \caption{Extraction of threshold voltage for a long-channel MOSFET from transconductance-to-current ratio (TCR) covering linear to saturation modes. (a) $G_mU_T/I_D$ obtained from $I_D$ vs. $V_G$ characteristics measured at different values of $V_{DS}$ (long-channel n-MOSFET) together with model \eqref{mosgmid}. (b) Criterion \eqref{mosgmidvth} for threshold voltage $nG_mU_T/I_D$ varies among two asymptotic values in linear and saturation modes. (c) Extracted threshold voltage as a function of $V_{DS}$ (as measured in (a)), compared to TCAD-simulated case without and with halo doping.}
	\label{vt_vds} 
\end{figure*}
\begin{equation}
\begin{split}
     %q_{s,d}=\frac{1}{2}L_W \left( 2e^{v_p-v_{s,d}} \right)
     q_{s,d}=(1/2)L_W \left( 2e^{v_p-v_{s,d}} \right)
      \label{qm}
\end{split}      
\end{equation}
where $L_W$ is the Lambert W (product logarithm) function. An analytical approximation with sufficient accuracy is \cite{lw} \\ $L_W(z) \cong \ln (1 + z)\left[ 1 - \ln (1 + \ln (1 + z))/(2 + \ln (1 + z))\right]$.      

Using the definition of threshold voltage in \cite{acc}, \cite{gcc} whereas threshold voltage is defined as the gate voltage for which the minimum potential in the channel ($v_s$) becomes equal to the pinch-off voltage $v_p$. Utilizing $v_p=v_s$ in \eqref{qm} gives,
\begin{equation}
\begin{split}
     %q_{s}&=\frac{1}{2}L_W \left( 2\right) \approx 0.4263 \\
     %q_{d}&=\frac{1}{2}L_W \left( 2e^{-v_{ds}} \right) \\
     q_{s}&=(1/2)L_W \left( 2\right) \approx 0.4263 \\
     q_{d}&=(1/2)L_W \left( 2e^{-v_{ds}} \right)
      \label{qmth}
\end{split}      
\end{equation}
where $v_{ds}=V_{DS}/U_T$.
Using the general charge-based expression for the transconductance-to-current ratio for MOSFETs,
\begin{equation}
\begin{split}
     \frac{G_{m}U_T}{I_D}=\frac{g_{m}}{i_d}=\frac{1}{n\cdot(1+q_s+q_d)}
      \label{mosgmid}
\end{split}      
\end{equation}
where $n$ is the weak inversion slope factor, evaluated from $n = \left[ max \left( G_mU_T/I_D \right)\right]^{-1}$ in weak inversion. Utilizing \eqref{qmth} we obtain $g_m/i_d$ at threshold,
\begin{equation}
\begin{split}
     \left.\frac{g_{m}}{i_d}\right|_{v_t}=\frac{1}{n\cdot(1.4263+\frac{1}{2}L_W \left( 2e^{-v_{ds}} \right))}
      \label{mosgmidvth}
\end{split}      
\end{equation}

Hence \eqref{mosgmidvth} provides an explicit, analytical criterion to determine threshold voltage over the full range of $V_{DS}$, that does not require any numerical iterations. In particular, for $v_{ds}=0$ in linear region  \eqref{mosgmidvth} gives $(1/2)L_W \left( 2e^{0}\right)=0.4263$,
\begin{equation} 
\begin{split}
\left.\frac{g_{m}}{i_d} \right|_{vt,lin} \approx \frac{0.54}{n}    \label{gmidthlin} 
\end{split}
\end{equation}
while for saturation, $(1/2)L_W \left( 2e^{-\infty} \right) \approx 0$,
\begin{equation} 
\begin{split}
\left.\frac{g_{m}}{i_d} \right|_{vt,sat} \approx \frac{0.7}{n}    \label{gmidthsat} 
\end{split}
\end{equation}

Hence, the threshold voltage is determined as the gate voltage at which $G_mU_T/I_D$ corresponds to a certain fraction of its maximum value in weak inversion, depending solely on $V_{DS}$. The extraction takes place in moderate inversion so series resistance and mobility degradation have minimum effect on extracted threshold voltage. 

%This methodology is applicable even if the threshold voltage definition differs from $v_p=v_s$. Considering as threshold voltage, the gate voltage for which the drift and diffusion source current components are equated gives $q_{s}=1$ and $v_p=2+v_s$ \cite{gcc}. Applying it to \eqref{qm} we obtain,
%\begin{equation}
%\begin{split}
%  \left.   \frac{g_{m}}{i_d}\right|_{vt}=\frac{1}{n\cdot(2+\frac{1}{2}L_W ( 2e^{2-v_{ds}}))}
%      \label{mosgmid2}
%\end{split}      
%\end{equation}

%The linear and saturation asymptotes of \eqref{mosgmid2} are $0.33/n$ and $0.5/n$ \cite{gcc}, %respectively. 

\section{Experiment and Discussion}

To validate our method, TCAD simulation and measurements are utilized. A long-channel ($L=10um$) NMOS device with oxide thickness $T_{ox}=3.45\,nm$ and uniform substrate doping $N_{sub}=10^{17}\,cm^{-3}$ was simulated with Silvaco Atlas, at room temperature, using CVT mobility and Shockley-Read-Hall (SRH) recombination models. Two cases without and with pocket/halo implants are considered, where the halo implants are uniformly doped areas ($2\cdot10^{18}cm^{-3}$) placed below drain and source extensions.

Transconductance-to-current ratio for the whole range of drain voltage, from $V_{DS}=10mV$ to $1.5V$, measured from a long-channel ($L=10um$) NMOS transistor of a $110nm$ CMOS technology is shown in Fig. \ref{vt_vds}(a). To confirm that the extracted $V_T$ is compatible with the used charge-based approach, $g_m/i_d$ is evaluated from \eqref{mosgmid} and demonstrated in Fig.\ref{vt_vds}(a). The normalized charges $q_s$ and $q_d$ in \eqref{mosgmid} are calculated using \eqref{qm} where $v_p$ is evaluated from \eqref{vpvg} using the extracted values of $V_T$ and $n$. The model fits measured characteristics adequately, confirming the consistency of procedure and model. 
The transconductance-to current-ratio $n \cdot G_mU_T/I_D$ which corresponds to $V_T$ vs. $V_{DS}$ and its asymptotic behavior in linear and saturation modes according to \eqref{gmidthlin} and \eqref{gmidthsat} is shown in Fig. \ref{vt_vds}(b).
The extracted threshold voltage $V_T$ vs. $V_{DS}$ is shown in Fig. \ref{vt_vds}(c). The latter also shows results of TCAD simulations. The case without halo implant features an almost invariable threshold voltage \cite{acc}, within $\pm 0.4mV$. The case with halo implant shows a characteristic increase of $V_T$ at $V_{DS} < 4\cdot U_T$ when the device is in non-saturation. The same qualitative increase observed in the real device may therefore clearly be attributed to halo implant. 
\begin{figure*}[t!]
	\centerline{%
		\begin{tabular}{ccc}
	   {\includegraphics[scale=0.12]{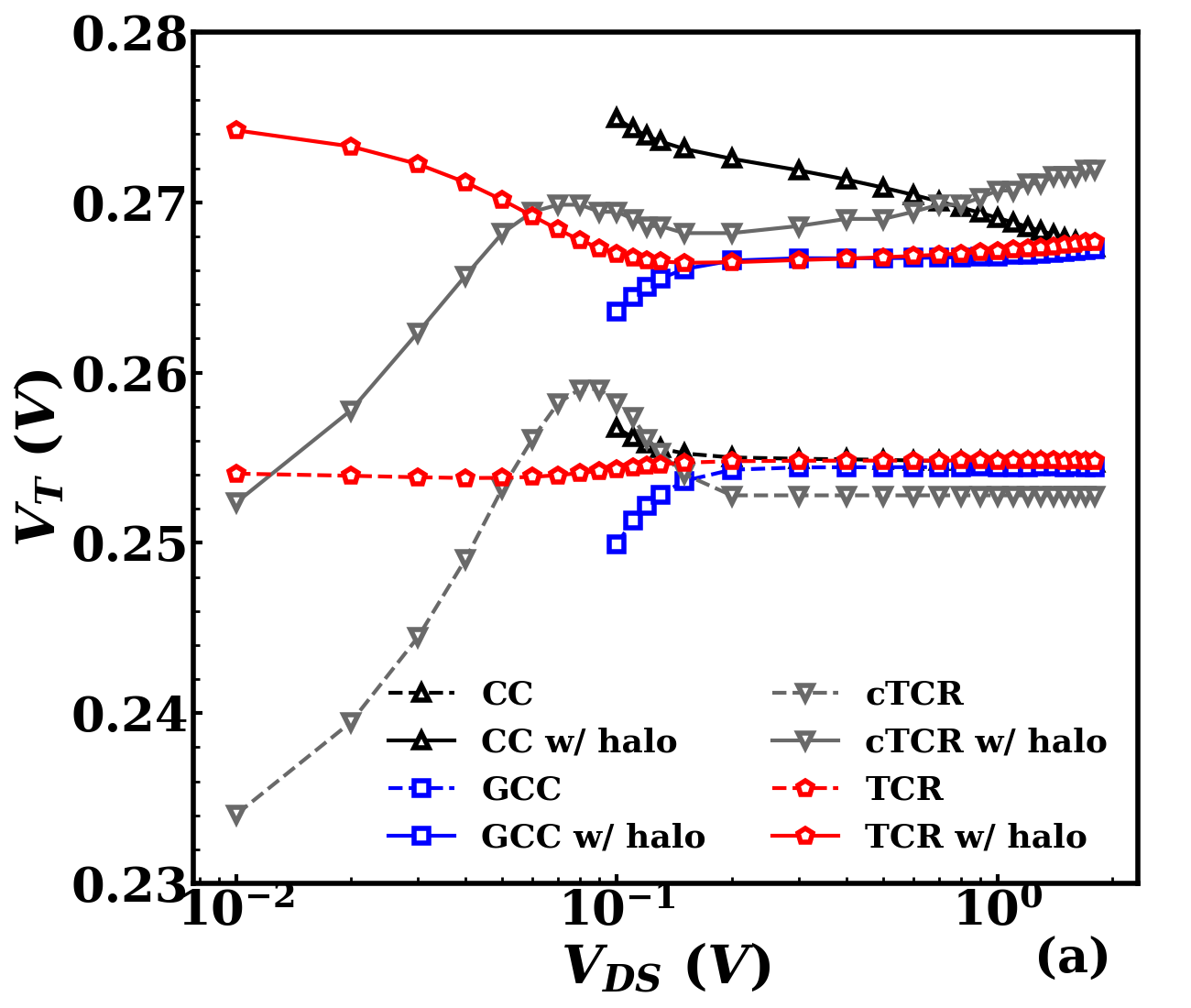}} &
	   {\includegraphics[scale=0.12]{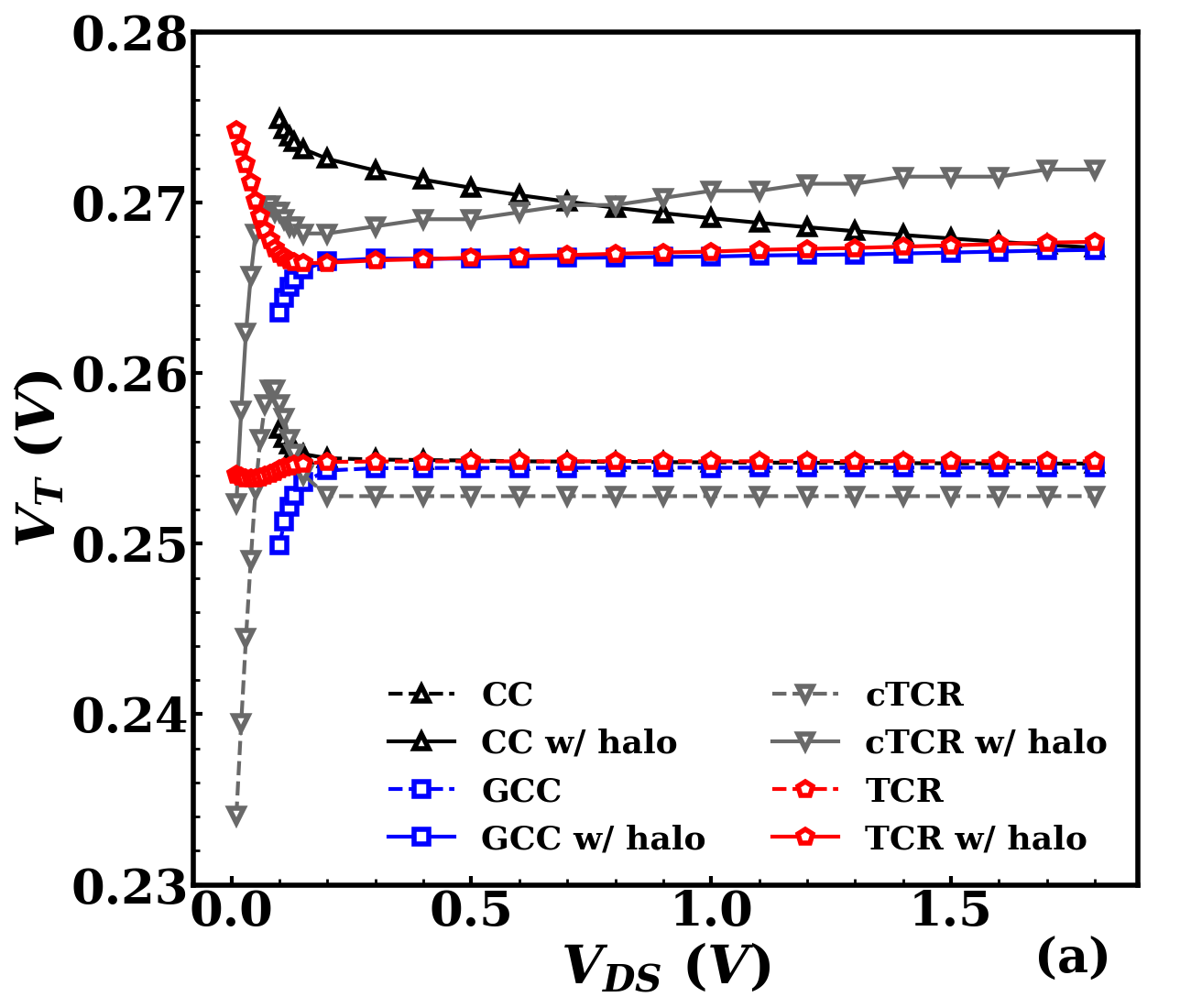}} &
	   {\includegraphics[scale=0.12]{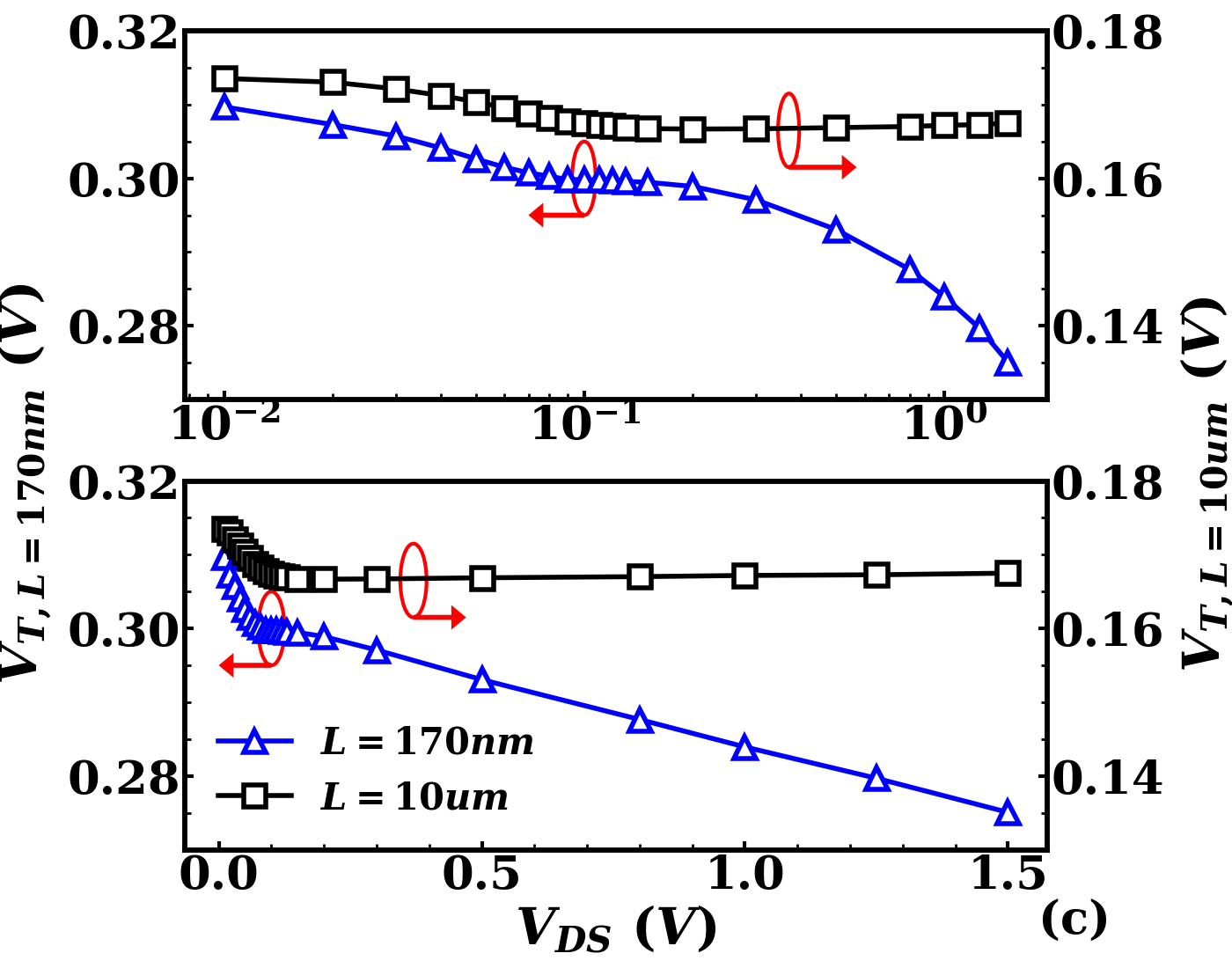}} 	   
	\end{tabular}} \caption{Extracted threshold voltage for  (a)-(b) TCAD-simulated MOSFET with and without halo implants utilizing constant-current (CC), generalized constant-current (GCC), peak of $\partial(G_m/I_{D})/\partial V_{GS}$ (cTCR) and this work (TCR).  (c) Long- and short-channel n-MOSFETs ($W=10um$) measured from 110nm CMOS process with pocket/halo doping (TCR). The characteristic increase of $V_T$ at low $V_{DS}$ is apparent in both long- and short-channel devices.} 
	\label{vtvdsall} 
\end{figure*}
%\begin{figure}[ht!]
%	\centerline{%
%		\begin{tabular}{ccc}
%	   {\includegraphics[scale=0.15]{GRAPHS/tcad/shortlong.png}} 
%	\end{tabular}} \caption{Extraction of threshold voltage for long- and short-channel ($L=1um, %170nm$) n-MOSFETs with pocket/halo doping from $G_m/I_D$ covering linear to saturation modes. The %characteristic increase of $V_T$ at low $V_{DS}$ is apparent in both long- and short-channel devices. %}
%	\label{vt_vds_sl} 
%\end{figure}

Extraction methods are compared in Fig. \ref{vtvdsall}(a)-(b) for the long-channel TCAD-simulated device with and without halo implant. For the CC method used here, the current criterion to extract $V_{T,CC}$ is $I_{T,sat}=0.618 \cdot I_0 \cdot (W/L)$ with technology current $I_0=605nA \,\,(621nA)$ in halo (non-halo) cases. $I_{spec}=I_0 \cdot (W/L)$ is the specific current obtained as in \cite{acc}. For the GCC method, the threshold criterion $V_{T,GCC}$ is determined as $V_G|_{V_P=0}$, corresponding to an inversion coefficient $IC=I_D/I_{spec}=0.608$ i.e. $n g_m/i_d=0.7$ \cite{gcc}. The procedure is repeated for each value of $V_{ds}$. Both CC and GCC methods are implemented for saturation conditions (i.e. $V_{ds}>4 \cdot U_T$ in moderate inversion). In the cTCR \cite{dgmi2010} method, $V_{T,cTCR}$ is the gate voltage corresponding to the peak of $\partial(G_m/I_D)/\partial V_G$. To obtain the needed resolution, we use a polynomial fitting of the data near the peak. The cTCR and this work's TCR methods provide $V_T$ in the full range of $V_{DS}$. The different behavior of the extracted threshold voltage is immediately apparent. For the non-halo device, $V_{T,cTCR}$ shows variation when departing from saturation, compared to the almost constant $V_{T,TCR}$, while the CC and GCC methods in saturation yield practically the same $V_T$ as cTCR and TCR methods (to within $2mV$). For the case with halo implants, the different behavior becomes more apparent due to the nature of each method. A common point to CC, GCC and TCR methods is the same estimation of change in $V_T$ at highest $V_{DS}$ among halo and non-halo cases.    

In Fig. \ref{vtvdsall}(c), the data of the long-channel device of Fig. \ref{vt_vds} are compared to a short-channel device of the same $110nm$ CMOS process. The short-channel device shows the same qualitative increase of $V_T$ at low $V_{DS}$, while exhibiting a clear drain-induced barrier lowering (DIBL) at higher $V_{DS}$.

The present method is applicable to a wide range of purposes, from process characterization to extraction of parameters in charge-based compact models, such as EKV \cite{ekv3}, ACM \cite{acm_cicc}, or BSIM-bulk \cite{bsim6vt} models, but may be easily adapted for use in other compact models. The method is based on the $G_m/I_D$ characteristic in moderate inversion, which is universal in many ways among different geometry, device types, temperature, and to some extent even for different types of FETs. A distinct feature is its applicability over the full range of $V_{DS}$ combined with moderate inversion operation, making it highly suitable for the investigation of DIBL and long-channel DIBL \cite{ldibl} effects.

\section{Conclusion}

A novel, fully analytical method for the extraction of threshold voltage in bulk MOSFETs over the full range of drain voltage has been presented. The method is based on the transconductance-to-current ratio $G_m/I_D$ in moderate inversion, and hence benefits from a low impact of high-field effects and series resistance. The charge-based modeling approach provides an explicit criterion for extracting $V_T$ directly from $G_m/I_D$ consistently for all $V_{DS}$. A practically constant value of threshold voltage $V_T$ is obtained in the full range of $V_{DS}$ from linear to saturation modes for ideal, long-channel devices. Halo-implanted devices show a characteristic increase of threshold voltage at low drain voltage, which is attributed to pocket/halo doping. The method is furthermore applicable to investigate DIBL effects.   

%\pagebreak

\end{document}